\documentclass[]{jfm}

\usepackage{graphicx}
\usepackage{graphbox}
\usepackage{newtxtext}
\usepackage{newtxmath}
\usepackage{mathtools}
\usepackage{bm}

\usepackage{natbib}
\usepackage{hyperref}
\hypersetup{
    colorlinks = true,
    urlcolor   = blue,
    citecolor  = black,
}

\usepackage{multirow}
\usepackage{accents}
\usepackage{comment}
\usepackage{subcaption}
\usepackage{xcolor} 
\usepackage{epstopdf, epsfig}
\usepackage{tikz}

\newcommand{\RomanNumeralCaps}[1]
\nolinenumbers
\graphicspath{{figures/}}
\usepackage{cleveref}

\newcommand{\Wi}{\ensuremath{\mathrm{Wi}}}

\shorttitle{Essential structure of `arrowhead' travelling waves}

\shortauthor{Zhu \& Kerswell}

\title{On the essential structure of exact traveling-wave solutions in viscoelastic flow}

\author{Lu Zhu\aff{1} \corresp{\email{lz447@cam.ac.uk}}, Rich. R. Kerswell\aff{1}}

 \affiliation{\aff{1}Department of Applied Mathematics and Theoretical Physics, University of Cambridge, Wilberforce Road, Cambridge CB3 0WA, UK}

\begin{document}

\maketitle

\begin{abstract}

We examine elastic travelling-wave (‘arrowhead’) solutions in a viscoelastic, unidirectionally body-forced flow, focusing on their existence and morphological changes as the Weissenberg number, \Wi,  and streamwise duct length, $L$,  are varied.
We find  that branch topology varies from an isola at low $L$ through a two-sided reconnection at intermediate $L$ to a branch which exists at asymptotically large Wi for larger $L$. At intermediate $L$ more than two arrowhead solutions can coexist at a given (\Wi, $L$) choice due to extra saddle node bifurcations. 
Secondly, the canonical arrowhead consists of two legs joined by an arched head that blocks throughflow and traps a counter-rotating vortex pair, while a polymer strand can emerge as a by-product of a strong extensional region attached/detached to the arrowhead arch. 
Thirdly, a minimal domain length $L_{\min}$ required to sustain an arrowhead is found to vary non-monotonically with $\Wi$; for $\Wi\ge 20$, detached-strand states control $L_{\min}$ with a relation $L_{\min}\approx 0.125\Wi+1.5$. 
And fourthly, in sufficiently long domains, the upper branch becomes a localised single arrowhead whose streamwise extent depends on $\Wi$, whereas the lower branch can  proliferate into a train of arrowheads at high $\Wi$, a phenomenon not previously reported. 

\end{abstract}


\section{Introduction}

The addition of a  small amount of polymer to a Newtonian fluid  can significantly alter its flow behaviour, producing distinct elastic phenomena \citep{steinberg2021elastic}. Such states have been explored across different configurations for drag reduction \citep{virk1975drag,white2008mechanics} and for enhancing turbulence, heat transfer, and mass exchange \citep[{e.g.}][]{qin2019upstream,kurzthaler2021geometric}. Among these elastic responses, the “arrowhead” elastic travelling wave -- first reported by \citet{page2020exact,dubief2022first} and subsequently observed over broad parameter space and geometries \citep{morozov2022coherent, beneitez2024multistability,zhu2024early,lellep2024purely} -- has attracted considerable attention~\citep[{e.g.}][]{datta2022,dubief2023elasto}. 

\begin{figure}
      \centering    
      \includegraphics[width=0.75\linewidth, trim=0mm 0mm 0mm 0mm, clip]{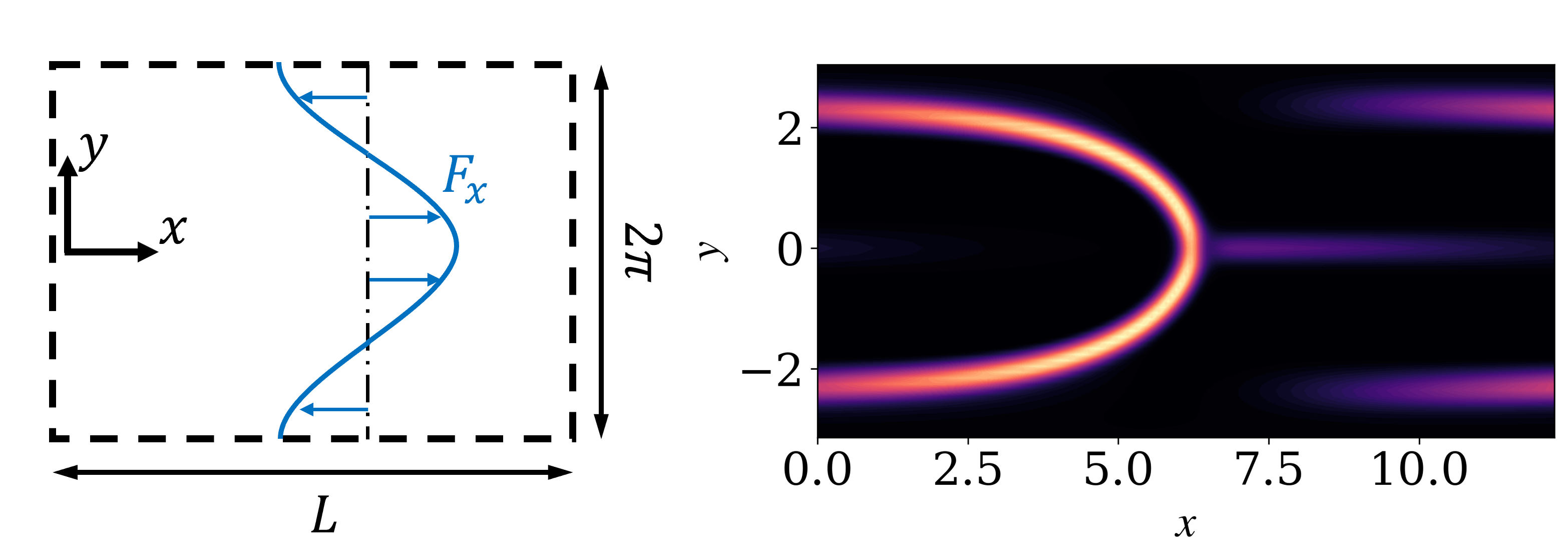}
    \caption{A schematic of Kolmogorov flow (a) and a typical arrowhead travelling wave structure (with a weak attached spike) (b).}
    \label{fig:schematic}
\end{figure}

The arrowhead solution is linked to the recently-identified centre–mode instability~\citep{garg2018viscoelastic, khalid2021continuous} in rectilinear viscoelastic flows that was long thought to be linearly stable \citep{larson1992instabilities}. \citet{page2020exact} demonstrated the subcritical nature of this linear instability and discovered a nonlinear travelling–wave solution in viscoelastic channel flow with an “arrowhead" appearance {on the upper branch of the bifurcated solutions}.
The canonical arrowhead consists of two extending polymer–stress bands joined by an arch - see Figure~\ref{fig:schematic}(b). In DNS, \citet{morozov2022coherent} reported an additional polymer “strand” attached to the arch and argued that it plays a critical role in sustaining the structure. However, the strand is not universal and can be absent in related flow (e.g. \citealp{buza2022finite}). Moreover, \citet{lewy2025revisiting} found that two arrowhead types can be triggered by different initial condition -- one has a clear head-strand and one hasn't. The origin of these states, and the conditions for the appearance of the strand, remain poorly understood.

Arrowhead morphology is also influenced by domain length, but this effect remains largely unexplored. 
Two questions are outstanding: (i) what is the minimal streamwise length of a (computational) periodic domain required to sustain an arrowhead, and (ii) under what conditions can the arrowhead localize? The first is central to understanding self-sustainment and criticality. Although domain effects have been noted in passing (e.g. \citealp{morozov2025narwhals}), a systematic study is lacking. Many works simply choose a domain commensurate with the centre–mode {instability} (e.g. \citet{buza2022finite}), despite the arrowhead being subcritical. The second question is related to the persistence of arrowheads in realistic, nonperiodic settings. Recent DNS \citep{morozov2025narwhals} provides partial evidence, but DNS can only access stable (typically upper-branch) states, leaving the full bifurcation structure and the conditions for localization underdetermined. 

To understand arrowhead behaviour, we use branch continuation  to track existence and topology changes in $(\Wi,L)$ parameter space with Wi being the Weissenberg number and $L$ the streamwise domain length. Continuation exposes hidden branches and clarifies transitions between multiple solution families. \citet{page2020exact} first applied continuation to the arrowhead in channel, implementing a Newton–Krylov (GMRES) scheme wrapped around a DNS time–stepper. Subsequently, \citet{buza2022finite} adopted a more computationally efficient (but {also more} memory–intensive) approach that solves the steady travelling–wave equations directly via a large algebraic system, identifying both an upper branch and a weaker lower branch over wide ranges of $\Wi$ and $\Rey$. However, 
those studies were restricted to relatively short domains. Here we also solve the large algebraic system, but in a doubly periodic Kolmogorov flow, which avoids the additional resolution required to resolve wall boundary layers in channel–flow settings.

The remainder of the paper is organized as follows. \S\ref{sec:method} summarizes the problem formulation and the continuation procedure. \S\ref{sec:results} presents the results in three parts: multiplicity of arrowhead states (\S\ref{sec:multiplicity}), presence and role of strands (\S\ref{sec:strand}), and the impact of domain length on solution transformations (\S\ref{sec:lx}). Final conclusions are given in \S\ref{sec:ccl}.

\section{Methodology}\label{sec:method}

\subsection{Governing equations}\label{sec:equations}

In this study we investigate arrowhead travelling-wave solutions (AHs) in a two-dimensional viscoelastic, unidirectionally body-forced (Kolmogorov) flow of Oldroyd–B fluids (Figure~\ref{fig:schematic}(a)), using direct numerical simulation (DNS) complemented by numerical branch continuation to track solution families. 
The streamwise ($x$) and cross-shear ($y$) directions are periodic while the body-force is applied to $x$-direction.
Following \citet{lewy2025revisiting}, we nondimensionalise velocities by the laminar peak velocity $\mathcal{U}$ and lengths by the $\mathcal{L}=L_f/2\pi$ ($L_f$ is the forcing wavelength).  
The nondimensional governing equations for the perturbation fields of the conformation tensor $\bm{\alpha}=(\alpha_{xx},\alpha_{xy};\alpha_{xy},\alpha_{yy})$, velocity $\bm{u}=(u,v)$, and pressure $p$ are
\begin{eqnarray}
    \label{eq:ns:mass}%
    \bm{\nabla} \cdot \bm{u} &=& 0,%
    \\
    \label{eq:ns:mom}%
    \frac{\partial \bm{u}}{\partial t} + \bm{u} \cdot \bm{\nabla} \bm{u} + \bm{U} \cdot \bm{\nabla} \bm{u} +\bm{u} \cdot \bm{\nabla} \bm{U} &=&
    - \bm{\nabla}p + \frac{\beta}{\Rey} \nabla^{2}\bm{u} + \frac{1-\beta}{\Rey\Wi}\bm{\nabla}\cdot\bm{\alpha}
    \\
    \label{eq:ns:oldb}%
    \frac{ \partial\bm{\alpha}}{\partial t} + \bm{u}\cdot\bm{\nabla}\bm{\alpha} + \bm{u}\cdot\bm{\nabla}\bm{A} +\bm{U}\cdot\bm{\nabla}\bm{\alpha} - \mathrm{DEF} - \mathrm{DEF}^T &=& -\frac{\bm{\alpha}}{\Wi}
    +\epsilon\nabla^2\bm{\alpha}
\end{eqnarray}
where the polymer deformation $\mathrm{DEF}=\bm{\alpha}\cdot\bm{\nabla}\bm{u} + \bm{\alpha}\cdot\bm{\nabla}\bm{U} +\bm{A}\cdot\bm{\nabla}\bm{u}$.
Dimensional parameters are Reynolds number $\Rey\equiv\mathcal{U} \mathcal{L}/\nu$ ($\nu$ is total kinematic viscosity),  Weissenberg number $\Wi\equiv \lambda\mathcal{U}/\mathcal{L}$ ($\lambda$ is polymer relaxation time), and viscosity ratio $\beta\equiv\nu_s/\nu$ ($\nu_s$ is solvent viscosity). Numerical simulation is performed around the viscoelastic laminar base
\begin{align}
\bm{U}(y) = \begin{pmatrix} \cos y \\ 0 \end{pmatrix},\qquad
\bm{A}(y) =
\begin{pmatrix}
1+\frac{\Wi^{2}}{1+\epsilon\,\Wi}\!\left(1-\frac{\cos(2y)}{1+4\epsilon\,\Wi}\right)
& -\frac{\Wi}{1+\epsilon\,\Wi}\,\sin y\\[6pt]
-\frac{\Wi}{1+\epsilon\,\Wi}\,\sin y
& 1
\label{base}
\end{pmatrix}.
\end{align}
A numerical diffusion term $\epsilon\nabla^2\bm{\alpha}$ is included to stabilize the simulation. The choice of polymer diffusion coefficient $\epsilon=10^{-3}$ is consistent with \citet{lewy2025revisiting}. 

\subsection{Traveling wave solution}\label{sec:tws}

The travelling wave solution (TWS) is computed by replacing the time derivatives $\partial/\partial t$ in Eqs.~(\ref{eq:ns:mom}) and (\ref{eq:ns:oldb}) by $-c\partial/\partial x$ ($c$ is the phase speed) and solving them with the pressure Poisson equation
\begin{equation}
    \nabla^2 p = -\bm{\nabla}\cdot (\bm{u} \cdot \bm{\nabla} \bm{u} + \bm{U} \cdot \bm{\nabla} \bm{u} +\bm{u} \cdot \bm{\nabla} \bm{U}) + \frac{1-\beta}{\Rey\Wi}\bm{\nabla}\cdot (\bm{\nabla}\cdot \bm{\alpha}).
    \label{eq:poisson}
\end{equation}
in place of Eq.~\ref{eq:ns:mass} (although this is used in deriving the Poisson equation). This unusual approach becomes competitive in a doubly-periodic domain and removes the phase/gauge ambiguity in $p$ as well as improving the conditioning of the Newton Jacobian.
In addition, an extra condition 
\begin{equation}
    \mathrm{Im}\{(\tilde{u}-\tilde{u}_0)(k_x=1,k_y=1)\}=0
    \label{eq:lux}
\end{equation}
is imposed ($\tilde{\cdot}$ denotes Fourier-space variables, $\tilde{u}_0$ is the initial guess). This removes the travelling wave phase degeneracy and thereby aids convergence~\citep[e.g.][]{buza2022finite}.

Numerically, this nonlinear elliptic equation system Eqs.~(\ref{eq:ns:mom}), (\ref{eq:ns:oldb}) (with $\partial/\partial t$ replaced), (\ref{eq:poisson}), and (\ref{eq:lux}) are solved in the 2D Fourier space using a direct method with a Newton-Raphson scheme to find TWS~\citep{buza2022finite}.
A steady DNS state or a previously converged TWS serves as the initial guess. Solutions are then continued along the bifurcation branch using pseudo-arclength continuation (see Appendix~A of \citet{buza2022finite}). 
Considering {the fact} that AHs are real and symmetric about the mid-plane $y=0$, only the non-negative Fourier modes ($k_x\geq 0, k_y\geq 0$) are retained, which greatly reduces memory usage.

\subsection{Numerical setup}\label{sec:numerics}

All numerical simulations are performed at a fixed $\Rey=0.5$ and $\beta=0.95$ representing an inertialess dilute polymeric solution. The ranges $\Wi\in(0,100)$ and channel streamwise length $L\in(\pi,50\pi)$ are searched to access multiple forms of AHs. The typical resolution is $\Delta x=\Delta y=2\pi/64$. For very large $L$, a coarser streamwise grid (e.g. $2\pi/32$) may be used. Grid-sensitivity checks indicate no noticeable impact on the results.

We use Dedalus~\citep{burns2020dedalus}, an open-source solver, to perform DNS of Eqs. (\ref{eq:ns:mass}-\ref{eq:ns:oldb}). It employs a Fourier–Fourier spectral scheme for spatial discretisation and a third-order, four-stage diagonally implicit–explicit Runge–Kutta scheme \citep{ascher1997implicit} for time stepping with a constant timestep $\Delta t=10^{-3}$.

\section{Results}\label{sec:results}
\subsection{Multiplicity of arrowhead states}\label{sec:multiplicity}

\begin{figure}
    \centering    
    \includegraphics[width=0.9\linewidth, trim=0mm 0mm 0mm 0mm, clip]{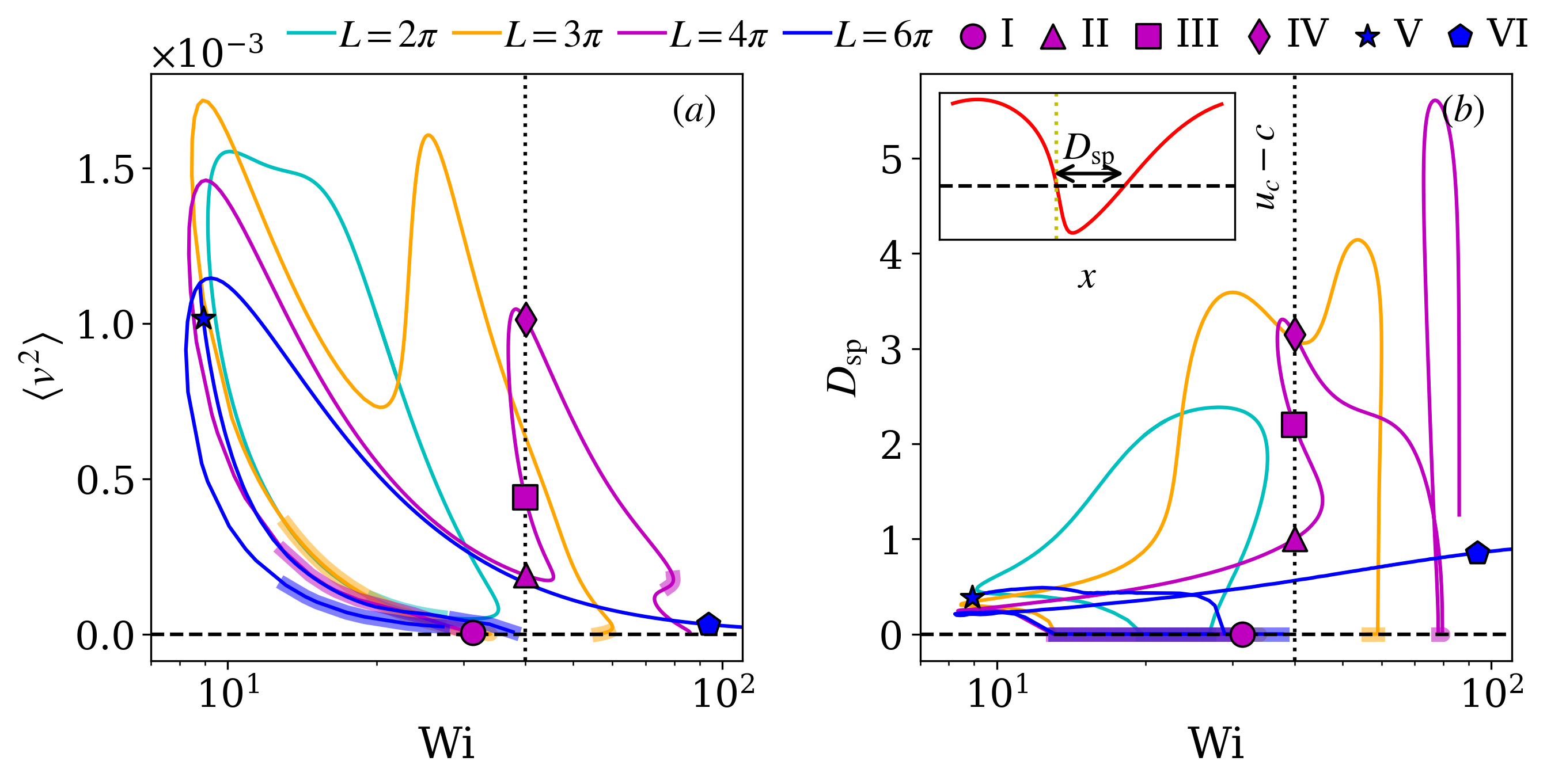}
    \caption{Branches of arrowhead solution continue in \Wi: (a) spanwise velocity fluctuations $\langle v^2 \rangle$ ($\langle\cdot\rangle$ denotes spatial averaging), (b) distance between the stagnation points, $D_{\mathrm{sp}}$. Thick lines indicate relative centerline velocity $u_c-c\geq 0$ throughout, corresponding to at most 1 stagnation point ($n_\mathrm{sp}\leq 1$). Insets: $u_{c}-c$ versus $x$. Dashed lines indicate $u_{c}-c=0$, and dotted lines mark the arch location.
    Flow fields of the 6 marked cases are shown in \cref{fig:flowfield}}
    \label{fig:branch}
\end{figure}

We begin by examining arrowhead branches continued in $\Wi$. Figure~\ref{fig:branch} shows that the branches exhibit distinct bifurcation structures and topological features at different streamwise length $L$.
At $L=2\pi$, the arrowhead branch forms an isola where the travelling wave exists only at finite amplitude and so does not bifurcate from the 1D laminar state (Eq.~\ref{base}). This also implies bistability, with two distinct stable solutions coexisting in parameter space.
For larger $L$, the branches reconnect to the laminar state at $\Wi \sim 30$, consistent with the emergence of a centre–mode instability \citep{lewy2025revisiting}.
At $L=4\pi$, multiple branches coexist. Around $\Wi =40$, these branches are linked by two saddle–node bifurcations, yielding three AHs at the same parameters -- two stable states separated by one unstable state (stability verified in DNS).
At the large–$\Wi$ limit, the $L=3\pi$ and $L=4\pi$ branches close and reconnect to the laminar state at finite $\Wi$. On the contrary, the $L=6\pi$ branch persists to arbitrarily large $\Wi$ and its amplitude asymptotes to zero as $\Wi\to\infty$. 
Comparable asymptotics occur for $L>6\pi$ (not shown), implying that the solution is effectively insensitive to $L$. Interestingly, the $L=3\pi$ and $L=6\pi$ solution curves connect in a modulational bifurcation shown by a blue star marker at $\Wi \approx 10$. In this, two wavelengths of a $L=3\pi$  arrowhead solution suffer a modulational instability to a $6\pi$ wavelength.

For a typical arrowhead, the centerline velocity $u_c$ decreases as the flow approaches the arch and increases afterwards. In a frame moving with the arrowhead, the relative centerline velocity $u_{c}-c$ can drop below zero, creating two stagnation points (see the inset of panel (b) of Figure~\ref{fig:branch}). Strong extensional flow develops around these stagnation points, which is implicated in the formation of AHs \citep{goffin2025follow,morozov2025narwhals}. 
However, this behaviour is not universal. Figure~\ref{fig:branch}(b) shows the distance $D_\mathrm{sp}$ between the two stagnation points in an arrowhead solution. In some cases, $u_{c}-c\geq 0$ is satisfied everywhere {which corresponds to $D_{\mathrm{sp}}=0$}. Such cases typically occur on the lower branch near the laminar state, where the structure is weak. Their persistence indicates that stagnation points may not play an essential role in sustaining AHs.
Apart from these cases, $D_{\mathrm{sp}}$ varies continuously, with substantial diversity across streamwise lengths $L$. For $L=2\pi$, $3\pi$, and $4\pi$, $D_{\mathrm{sp}}$ shows no clear asymptotic trend with $\Wi$. By contrast, for $L=6\pi$, once past the left saddle–node, $D_{\mathrm{sp}}$ increases asymptotically with $\Wi$.

\begin{figure}
      \centering    
      \includegraphics[width=0.99\linewidth, trim=0mm 0mm 0mm 0mm, clip]{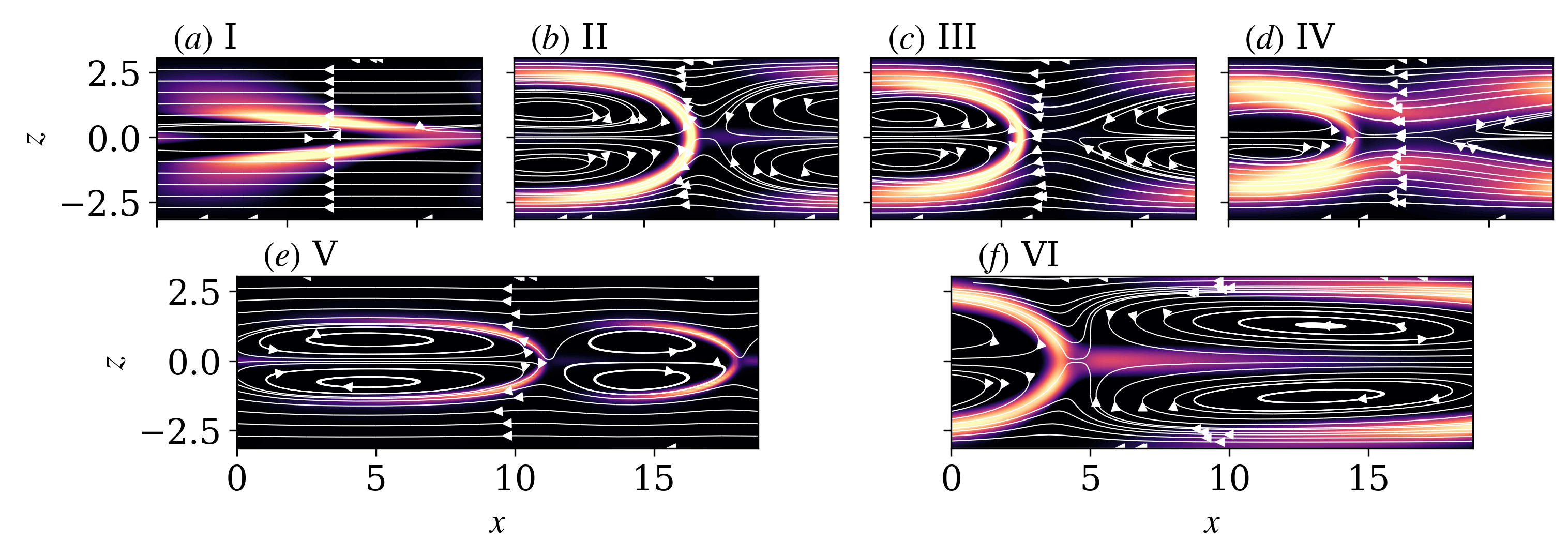}
    \caption{Typical structures of arrowheads: (a) I, near-onset arrowhead solution resembles to eigenfunction of centre-mode instability, (b) II, lower branch with attached strand, (c) intermediate unstable solution with detached strand (d) III, upper branch arrowhead with detached weak strand, (e) IV, modulated solution at $L=6\pi$, (f) upper asymptote solution. Contours denote the normalized trace of polymer conformation $\mathrm{tr}\bm{\alpha}/\max(\mathrm{tr}\bm{\alpha})$, lines indicate the streamlines on the coordinate moving with the phase speed. 
    }
    \label{fig:flowfield}
\end{figure}

In Figure~\ref{fig:flowfield} we show representative structures of AHs. Case~I lies close to the laminar state and closely resembles the eigenfunction of the centre–mode instability (see \citet{lewy2025revisiting}) consistent with the fact that arrowhead is the nonlinear saturation of the centre–mode instability.
In the moving frame of the AHs, AHs consist of two legs joined by an arched head that blocks throughflow and traps a downstream counter-rotating vortex pair.
Panels (b–d) display three solutions straddling a fold (saddle-node) bifurcation: II and IV are stable, whereas III is unstable. In Case~II a clear strand is attached to the arch while such strand is not distinguishable from the arch in Cases~III and IV. In fact, a detached, weak strand forms downstream within the counter-rotating vortex pair, likely generated by the intervening extensional corridor and unrelated to the {convex side of the downstream arch}. For larger $L$, the continuation diagram becomes richer, leading to a modulated arrowhead and weak nested arrowhead at extreme \Wi.

\subsection{Persistence of strand}\label{sec:strand}

%
%
\begin{figure}
    \centering    
    \includegraphics[width=0.99\linewidth, trim=0mm 0mm 0mm 0mm, clip]{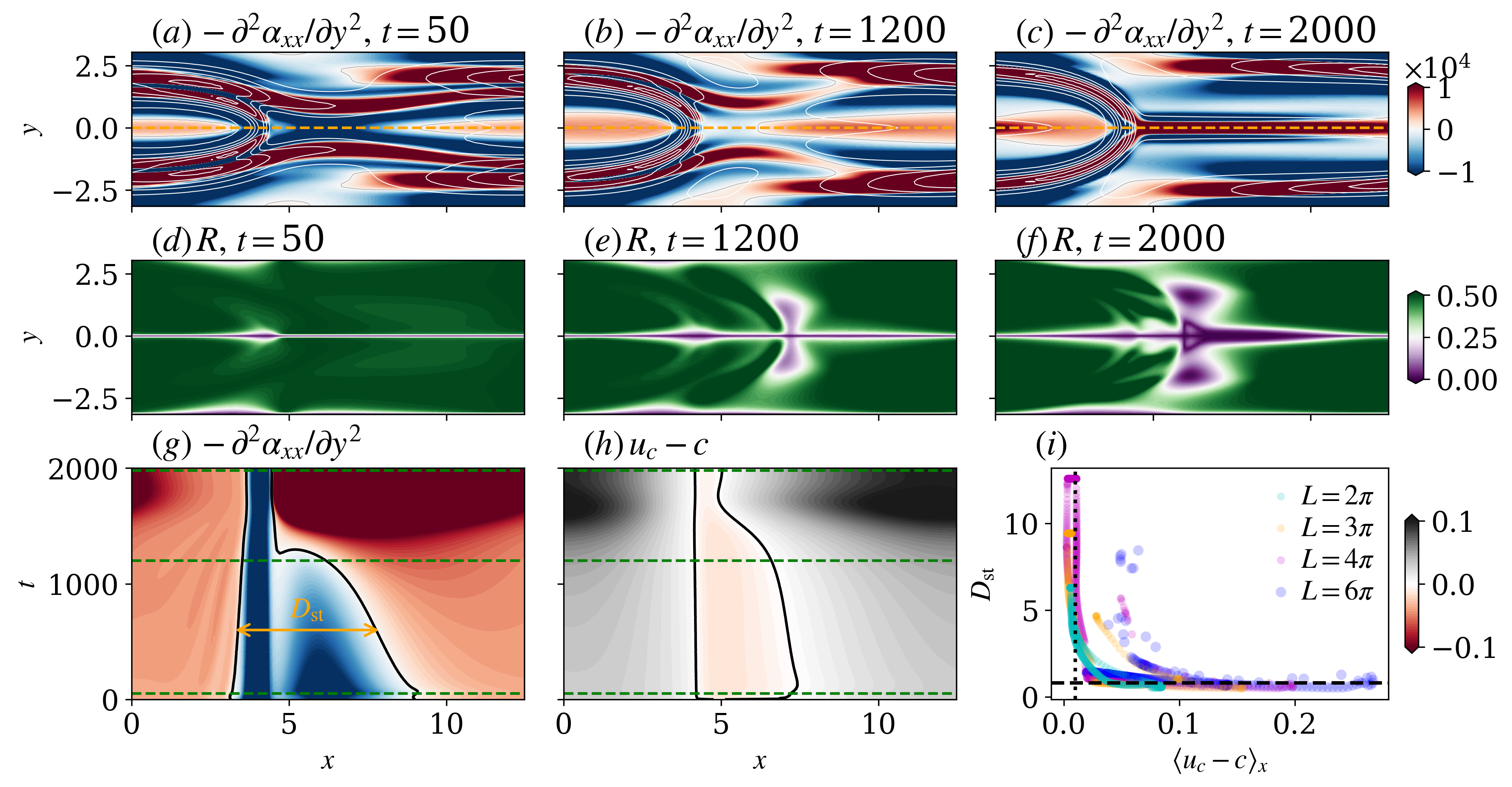}
    \caption{Transition between detached and attached strands in DNS: (a-c) instantaneous fields of strand measure $-\partial^2 \alpha_{xx}/\partial y^2$ (colors) and $\mathrm{tr}\bm{\alpha}$ (lines) at $t=50$, $1200$ and $2000$ and (d-f) measure of flow extension $R=|\Omega|/|S|$ at the same times, (g,h) Space-time $x-t$ diagrams of (g) $-\partial^2 \alpha_{xx}/\partial y^2$ and (h) $u_c-c$ at $y=0$. Green dashed lines indicate the times shown in panels (a–f). (i) the $x$-averaged relative centerline velocity $\langle u_c-c\rangle_x$ versus the arch-strand separation $D_\mathrm{st}$, measured as the distance between the two zero crossings of $-\partial^2 \alpha_{xx}/\partial y^2$ shown in (g).   }
    \label{fig:nose}
\end{figure}

In the literature, the persistence of an upstream “strand” (or “spike”) ahead of the arch is often reported and treated as an essential feature of AHs \citep{morozov2022coherent,morozov2025narwhals}, whereas {in other studies (e.g. Figure 7 in \citealp{buza2022finite}; Figure 10 in \citealp{lewy2025revisiting}; Figure~1(c) in \citealp{nichols2025period} )} such strands are not observed. In Figure~\ref{fig:flowfield} we have already shown the coexistence of three distinct solutions -- one with a clear strand (II) and two without (III, IV). To interrogate this, we perform a transient DNS at $L=4\pi$, evolving from a `no-strand' solution at $\Wi=38$ (near Case~IV) to a strand-bearing solution at $\Wi=36$ in Figure~\ref{fig:nose}. We use $-\partial^{2}\alpha_{xx}/\partial y^{2}$ as a strand indicator: positive values denote a convex polymer–stretch profile (i.e., a strand). In Figure~\ref{fig:nose}(a–c) and Movie I in supplementary, a weak strand is initially present between the downstream counter-rotating vortices, detached from the downstream arch. As $\Wi$ decreases, it extends and strengthens, ultimately attaching to the downstream arch and forming a pronounced strand. 
Figure~\ref{fig:nose}(d–f) shows an measure of flow extension, $R=\lVert \Omega\rVert_{2}/\lVert S\rVert_{2}$ ($\Omega$ and $S$ are the rotation and strain-rate tensors). In panel (d), extension is weak and only a weak strand forms within the extensional corridor between the counter-rotating vortices. By contrast, in panel (f), a strong extensional region develops near the arch, generating the attached strand.
The occasional absence of a strand indicates that it is \emph{not required} for the persistence of the arrowhead state. Rather, it is a by-product of extensional flow, which need not occur upstream of the arch.

In Figure~\ref{fig:nose}(g,h) we present space–time ($x$–$t$) diagrams of $-\partial^{2}\alpha_{xx}/\partial y^{2}$ and $u_{c}-c$ at $y=0$, respectively. The arch position at each time is aligned to isolate structural changes. In panel~(g), a strong, persistent negative band marks the arch and varies little over time. The upstream negative band, representing the strand–arch gap, shrinks gradually and suddenly disappears at $t\approx 1200$. Thereafter, a positive band (the strand) intensify and direct contact with the arch, indicating strand attachment. Consistently, panel (h) shows the evolution of $u_{c}-c$: the two zero crossings (black lines, stagnation points) approach one another. A subsequent jump to large positive values indicates enhanced flow extension in the attached state. Interestingly, the left stagnation line is vertically fixed and, in fact, as in Figure~\ref{fig:branch}(b), coincides with the arch head.

The strand attachment correlates with the relative centerline velocity. In Figure~\ref{fig:nose}(i) we plot the joint distribution of the arch–strand separation, $D_{\mathrm{sp}}$ (defined as the distance between the two zero crossings of $-\partial^{2}\alpha_{xx}/\partial y^{2}$; see panel (g)), and the $x$-averaged relative centerline velocity, $\langle u_{c}-c\rangle_x$, for all branches in Figure~\ref{fig:branch}. Many points cluster near a horizontal band at $D_{\mathrm{sp}}\approx 0.8$, indicating arrowheads with attached strands. {This separation
suggests the nearly uniform arch thickness of the attached-strand cases across different $\Wi$ and $L$.}
A second cluster appears along a vertical band at $\langle u_{c}-c\rangle_{x}\approx 0.01$, characteristic of arrowheads with a detached strand.
In practice, $\langle u_{c}-c\rangle_{x}$ can serve as a diagnostic: values near $0.01$ indicate a detached strand, whereas larger values suggest attached strands.

\subsection{Duct length and localisation}\label{sec:lx}

%
%
\begin{figure}
    \centering    
    \includegraphics[width=0.8\linewidth, trim=0mm 0mm 0mm 0mm, clip]{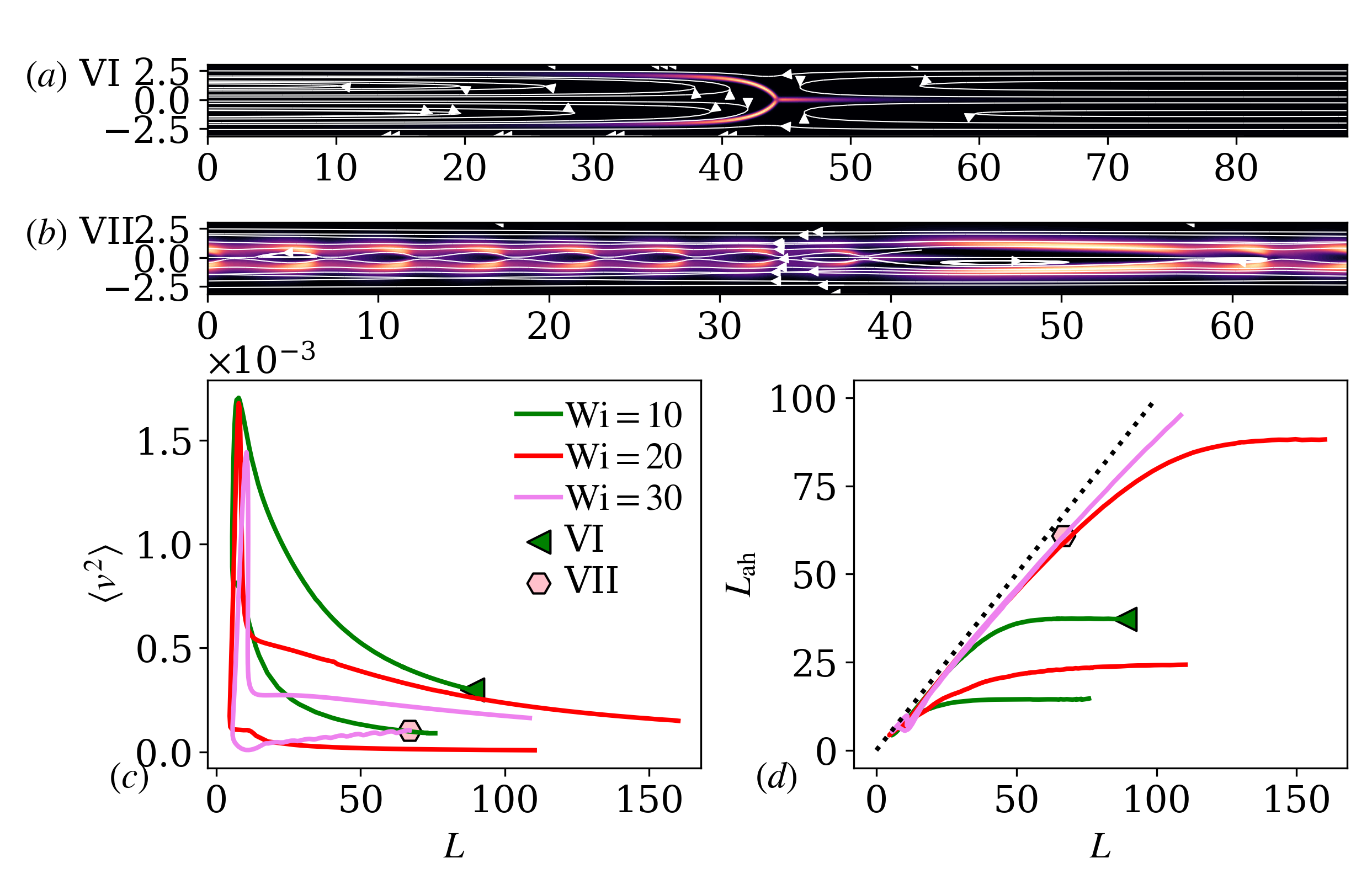}
    \caption{Influence of streamwise length: (a) localisation (VI) and (b) train of arrowheads in the long channel (colors: $\mathrm{tr}\bm{\alpha}/\max(\mathrm{tr}\bm{\alpha})$, lines: streamlines on the coordinate moving with $c$).  Branches of AHs projected onto (c) the $\langle v^2 \rangle$ vs streamise length $L$, and (d) the arrowhead length ($L_\mathrm{ah}$) vs $L$ space, respectively. The black dash line is $L_\mathrm{ah}=L$.  }
    \label{fig:lx_dep}
\end{figure}

In Figure~\ref{fig:branch}, we have shown that travelling-wave branches depend sensitively on the streamwise length $L$. To further clarify this dependence and identify possible spatial localization of AHs, we now perform branch continuation in $L$. We first define a localisation measure, $L_{\mathrm{ah}}$, as the streamwise length of domain containing $95\%$ of the total stretched polymers, quantified by

\begin{equation}
    {L_\mathrm{ah}=\int_L \bm{1}_{[\langle\mathrm{tr}\bm{\alpha}\rangle_{y}\geq \tau^\star]}\,dx, \quad \text{where} \,  \tau^\star=\mathrm{sup}\{\tau: \frac{\int_L\langle  \mathrm{tr}\bm{\alpha} \rangle_y\bm{1}_{[\langle  \mathrm{tr}\bm{\alpha} \rangle_y \geq\tau]}dx}{\int_L\langle  \mathrm{tr}\bm{\alpha} \rangle_y\,dx}\geq 95\% \},}
\end{equation}
{where $\langle \cdot \rangle_y$ denotes averaging over $y$.}
This definition eliminates bias from variations in the overall $\mathrm{tr}\,\bm{\alpha}$ magnitude induced by changes in $\Wi$, $L$ or branch.

In Figure~\ref{fig:lx_dep}(c), the upper and lower branches are observed at $\Wi=10,\,20$ and $30$. The upper branch corresponds to stronger saturated arrowheads (cf. panel~(a)), whereas the lower branch is weaker. Notably, for $\Wi=30$ the lower branch experience a sequence of bifurcations that produces a \emph{train of arrowheads} (panel~(b)), visible as vibration in the continuation curve. Phenomenologically, additional arches proliferate as $L$ increases via destabilization and reconnection of the elongated arrowhead. To our knowledge, this phenomenon has not been previously reported.

In Figure~\ref{fig:lx_dep}(d) we investigate localization by comparing $L_{\mathrm{ah}}$ with $L$. When a branch aligns with the diagonal $L_{\mathrm{ah}}=L$, the state is effectively non-localized (small deviations is due to the 95\% threshold). As the branch curve bends away from the diagonal and plateau, the arrowhead localizes and further increasing $L$ does not change its length or shape.

\textit{Upper branch.} Localization depends on $\Wi$. For $\Wi=10$, $L_{\mathrm{ah}}$ plateaus at $\approx 37$ for $L\approx 57$; for $\Wi=20$, $L_{\mathrm{ah}}\approx 88$ at $L\approx 132$. For $\Wi=30$, no plateau is reached within the explored $L$, but one is expected at sufficiently large $L$.

\textit{Lower branch.} Plateaus occur earlier and at smaller $L_{\mathrm{ah}}$. For $\Wi=10$, a plateau appears at $L\approx 25$ with $L_{\mathrm{ah}}\approx 14$; for $\Wi=20$, at $L\approx 54$ with $L_{\mathrm{ah}}\approx 24$. Owing to the arrowhead train, the lower branch at $\Wi=30$ does not localize.

\begin{figure}
    \centering    
    \includegraphics[width=0.9\linewidth, trim=10mm 15mm 0mm 10mm, clip]{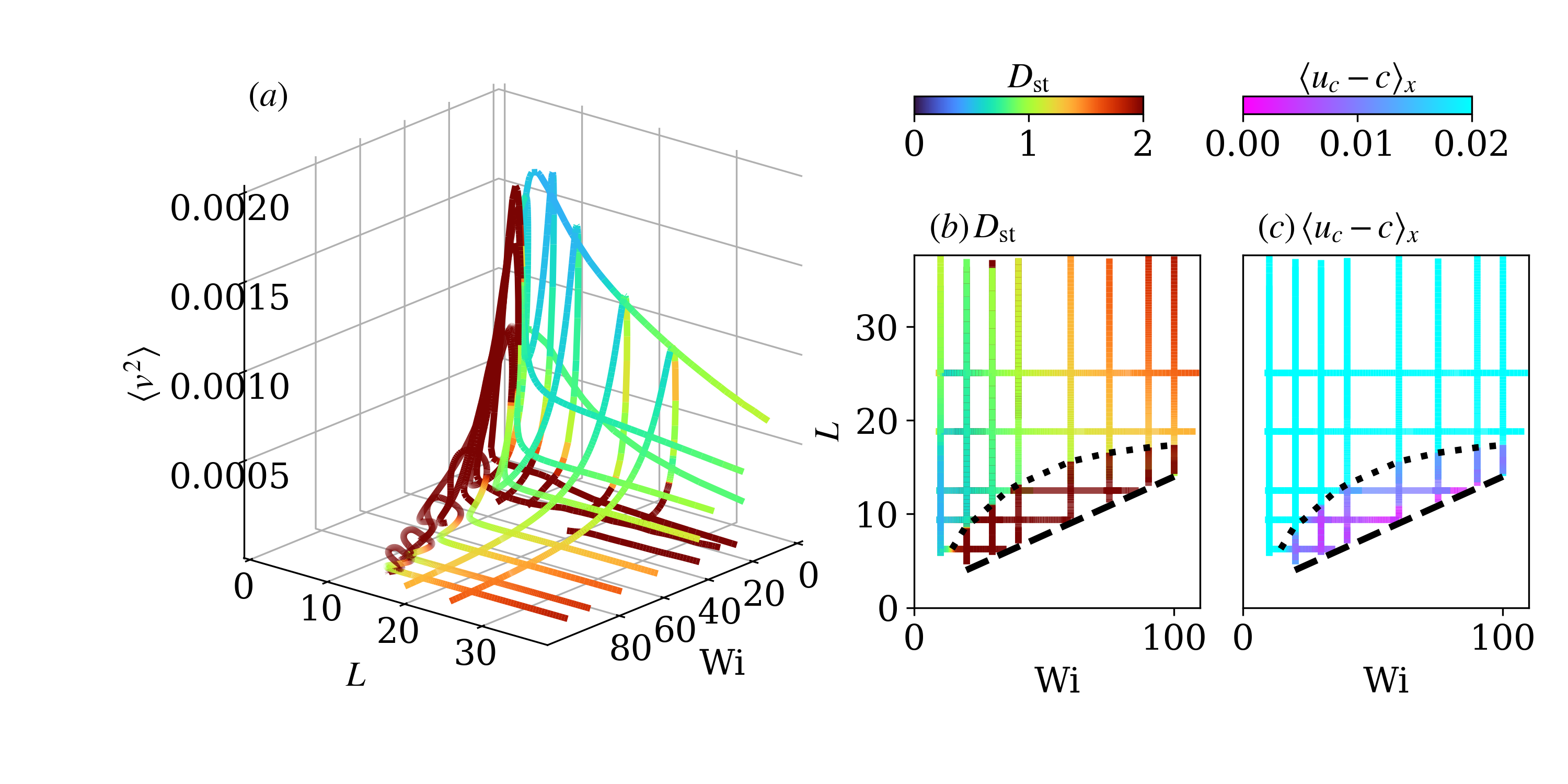}
    \caption{Continuation of arrowhead branches in $\Wi$ and $L$:(a) branches in 3D space ($\Wi,L,\langle v^2\rangle$) and (b,c) projection of these branches on the $\Wi-L$ space. }
    \label{fig:3d}

\end{figure}

To summarize the solution landscape, Figure~\ref{fig:3d}(a) presents a joint view of branches continued in $\Wi$ and $L$, plotted against $\langle v^{2}\rangle$. Notably, arrowhead intensity (measured by $\langle v^{2}\rangle$) is largest at relatively small $\Wi$ and short $L$, despite the states being elastically driven. Colors represent the arch–strand separation $D_{\mathrm{sp}}$. A distinct region of large $D_{\mathrm{sp}}$ appears for $\Wi>20$ and $L<6\pi$, indicating detached strands, a second detached cluster occurs on the lower branch near the laminar state.

In panel (b), the projection onto the $\Wi$–$L$ plane delineates the existence boundary of AHs. The minimum $\Wi$ varies little with $L$, whereas the minimum $L$ changes non-monotonically with $\Wi$: it decreases from $\Wi\approx 10$ to $\Wi\approx 20$, then increases approximately linearly for $20\lesssim \Wi \lesssim 100$, 
i.e. $L_{\min}\approx 0.125\,\Wi + 1.5$.  
The turnover coincides with the
attached$\rightarrow$detached transition (visible as a sharp color change), after which detached states determine $L_{\min}$.
This linear trend cannot persist indefinitely: for detached states, the admissible $L$ gradually approaches $6\pi$, beyond which only attached AHs remain, as the $L=6\pi$ branch tends to the laminar base as $\Wi\to\infty$. Consequently, we expect $L_{\min}\to 6\pi$ at sufficiently large $\Wi$ (estimated as $\approx 140$ by equating $6\pi$ with $0.125Wi+1.5$). 
Interestingly, panel (c) for the $x$-averaged centerline relative velocity, $\langle u_{c}-c\rangle_x$ shows a similar pattern: the detached arrowhead region coincides with small values of $\langle u_{c}-c\rangle_x$. This suggests again that $\langle u_{c}-c\rangle_x$ can serve as an effective diagnostic for distinguishing detached/attached AHs.

\section{Conclusions}\label{sec:ccl}

This study integrates travelling-wave branch continuation and DNS to examine the morphological features of the arrowhead solution under variations in Weissenberg number (\Wi) and streamwise length $L$ in a viscoelastic, unidirectionally body-forced flow.

In \S\ref{sec:multiplicity}, we showed that the branch topology evolves continuously with increasing $L$: from an isola, to a branch that reconnects with the laminar state on both sides, and finally to a branch that persists to arbitrarily large $\Wi$ with amplitude asymptoting to zero as $\Wi\to\infty$. This rich dynamics leads to multiple coexisting arrowhead solutions.

Diagnosing these solutions (\S\ref{sec:strand}) reveals the essential arrowhead structure: an arched head joined by two stretching legs that block the throughflow and trap a downstream counter-rotating vortex pair. A head-strand can emerge as a by-product of strong extensional regions but is not required for the persistence of AHs. This yields two geometric categories -- \emph{attached} vs \emph{detached} strands -- depending on whether the strand connect the arch. In practice, the mean relative centerline velocity $\langle u_{c}-c\rangle_x$ discriminates the two categories: detached cases cluster near $\langle u_{c}-c\rangle_x \approx 0.01$, whereas attached cases exhibit substantially larger values.

In \S\ref{sec:lx}, we continue the branches in $L$. In sufficiently long domain, the upper branch retains a single, classical arrowhead whose streamwise lengths depend on $\Wi$ and becomes localized at sufficient $L$. The lower branch also localizes at $\Wi=10$ and $20$, but undergoes bifurcations at $\Wi=30$ in which the legs reconnect to produce a train of arrowheads. In the short-domain limit, the minimal sustaining length depends non-monotonically on $\Wi$, due to the change in dominance between attached and detached states at the smallest $L$. For $\Wi\geq 20$, the detached state controls the bound and the minimal length follows $L_{\min}\approx 0.125\,\Wi + 1.5$.

\vspace{0.1cm}

Declaration of interest: The authors report no conflict of interest.

\bibliographystyle{jfm}
\bibliography{References_ve.bib}

\end{document}